\documentclass[conference]{IEEEtran}
\IEEEoverridecommandlockouts
\usepackage{cite}
\usepackage{amsmath,amssymb,amsfonts}
\usepackage{graphicx}
\usepackage{textcomp}
\usepackage{algorithm}
\usepackage{algpseudocode}

\usepackage[letterpaper, top=.75in, bottom=1in, left=0.625in, right=0.625in]{geometry}
\usepackage{xcolor}
\def\BibTeX{{\rm B\kern-.05em{\sc i\kern-.025em b}\kern-.08em
T\kern-.1667em\lower.7ex\hbox{E}\kern-.125emX}}
\begin{document}

\title{
Mitigating Pilot Contamination and Enabling IoT Scalability in Massive MIMO Systems\\

\thanks{This research was partly funded by Palmer Department Chair Endowment Fund at Iowa State University.}
}

\author{\IEEEauthorblockN{ Muhammad Kamran Saeed}
\IEEEauthorblockA{\textit{Department of Electrical and}\\{ Computer Engineering} \\
\textit{Iowa State University}\\
Ames, USA \\
Kamran@iastate.edu}
\and
\IEEEauthorblockN{Ahmed E. Kamal (late)}
\IEEEauthorblockA{\textit{Department of Electrical and}\\{ Computer Engineering} \\
\textit{Iowa State University}\\
Ames, USA \\
Kamal@iastate.edu}
\and
\IEEEauthorblockN{Ashfaq Khokhar}
\IEEEauthorblockA{\textit{Department of Electrical and}\\{ Computer Engineering} \\
\textit{Iowa State University}\\
Ames, USA \\
Ashfaq@iastate.edu}

}

\maketitle

\begin{abstract}
Massive MIMO is expected to play an important role in the development of 5G networks. This paper addresses the issue of pilot contamination and scalability in massive MIMO systems. The current practice of reusing orthogonal pilot sequences in adjacent cells leads to difficulty in differentiating incoming inter- and intra-cell pilot sequences. One possible solution is to increase the number of orthogonal pilot sequences, which results in dedicating more space of coherence block to pilot transmission than data transmission. This, in turn, also hinders the scalability of massive MIMO systems, particularly in accommodating a large number of IoT devices within a cell. To overcome these challenges, this paper devises an innovative pilot allocation scheme based on the data transfer patterns of IoT devices. The scheme assigns orthogonal pilot sequences to clusters of devices instead of individual devices, allowing multiple devices to utilize the same pilot for periodically transmitting data. Moreover, we formulate the pilot assignment problem as a graph coloring problem and use the max k-cut graph partitioning approach to overcome the pilot contamination in a multicell massive MIMO system. The proposed scheme significantly improves the spectral efficiency and enables the scalability of massive MIMO systems; for instance, by using ten orthogonal pilot sequences, we are able to accommodate 200 devices with only a 12.5\% omission rate.
\end{abstract}


\section{Introduction}

Massive MIMO enhances 5G networks, enabling greater spectral efficiency, capacity, coverage, and reliability. This technology also boosts energy efficiency, crucial for widespread IoT deployment. The combination of IoT devices and massive MIMO is expected to enable a range of new services and applications that will transform life experiences.

Massive MIMO enhances spectral efficiency by scheduling multiple devices for concurrent spectrum access. Feres et al. \cite{b3} proposed a scheduling scheme to cluster devices having minimal interference for effective resource sharing but without capping cluster size. Thus, high device volumes in clusters could lead to delays as devices await their turn to transmit, questioning the device's ability to transmit data effectively while meeting timing requirements. Therefore, we propose a scheduling scheme by taking into account data transmission time and time period, alongside introducing a novel pilot sequence allocation scheme to mitigate inter-cell interference. 
\par In massive MIMO, acquiring channel state information plays a key role in base stations communicating with multiple devices simultaneously and at the same frequency~\cite{b1}. The popular way to get the channel state information is by using orthogonal pilot signals. Orthogonal pilot signals are predefined symbols assigned to each device within a cell to estimate the channel between the base station and the device. However, the number of orthogonal pilot sequences are limited. Therefore, pilot sequences are usually reused in the adjacent cells. Since the pilot sequences are transmitted synchronously in all cells, the base station finds it difficult to differentiate whether the incoming pilot sequence comes from the adjacent or within a cell. This phenomenon is called inter-cell pilot contamination (PC) and is considered a  critical limiting factor in enhancing spectral efficiency\cite{b2}. We propose to utilize max K-cut graph partitioning method to overcome this issue.

\par To address intra-cell PC, it is essential to utilize orthogonal pilot sequences for each devices within a cell. However, this strategy encounters scalability issues due to the limited coherence block space, restricting the integration of multiple IoT devices. Consequently, the idea of periodic data transmission to a data concentrator by IoT devices came into consideration. Therefore, rather than giving each device a unique orthogonal pilot sequence, we allocate one to a cluster of devices, and each device can utilize the same pilot to transmit data in its turn. Simulation results show enhanced spectral efficiency and scalability of IoT devices in massive MIMO systems.

\section{Contribution}

The contributions of this paper are summarized as follows:
\begin{enumerate}

    \item To address the issue of scalability, we propose a clustering-based pilot assignment scheme to efficiently integrate more IoT devices in a massive MIMO system through the efficient use of scarce pilot signal resources.  
    \item We propose a modified Kfaster medoid clustering algorithm to cluster devices by considering spatial correlation among them.
    \item We address the inter-cell pilot contamination problem by formulating the original problem as a graph coloring problem and maximizing the throughput by smartly allocating pilot signal between clusters using the max K-cut graph partitioning approach. 
   
\end{enumerate}

\section{System Model}
This paper considers a multi-cell massive MIMO network with L cells. Each cell has $K$ IoT devices clustered in $C$ clusters, and each cluster contains $d$ devices represented by $U_d$. The base station is equipped with $M$ antennas. Moreover, a massive MIMO system is working in uplink data transmission. The channel between the $k_{th}$ IoT device of the ${i_{th}}$ cell, where ${i\in\{1,...,L\}}$, with the base station of the ${i_{th}}$ cell, is denoted by $h_{ik}^{i} \in \mathbb{C}^M$. The interference channel between the $k_{th}$ device, of the ${j_{th}}$ cell with the base station of the ${i_{th}}$ cell is denoted by $h_{jk}^{i} \in \mathbb{C}^M$. A tractable way to model the spatially correlated Rayleigh fading channel with no line of sight can be represented as $\mathbf {h}_{ik}^{j} \sim \mathcal {N}_{\mathbb {C}} \left ({\mathbf {0}_{M}, \mathbf {R}_{ik}^{j} }\right)$.

Where the ${R}_{ik}^{j} \in \mathbb{C}^{M \times M}$ represents a spatial correlation matrix. Eigen-structure of ${R}_{ik}^{j}$ captures the hidden spatial correlation properties of ${h}_{ik}^{j}$, with the assumption that certain spatial directions are statistically more likely to contain a signal component than others.      

\begin{figure}[htp]
    \centering
    \includegraphics[width=8.65cm]{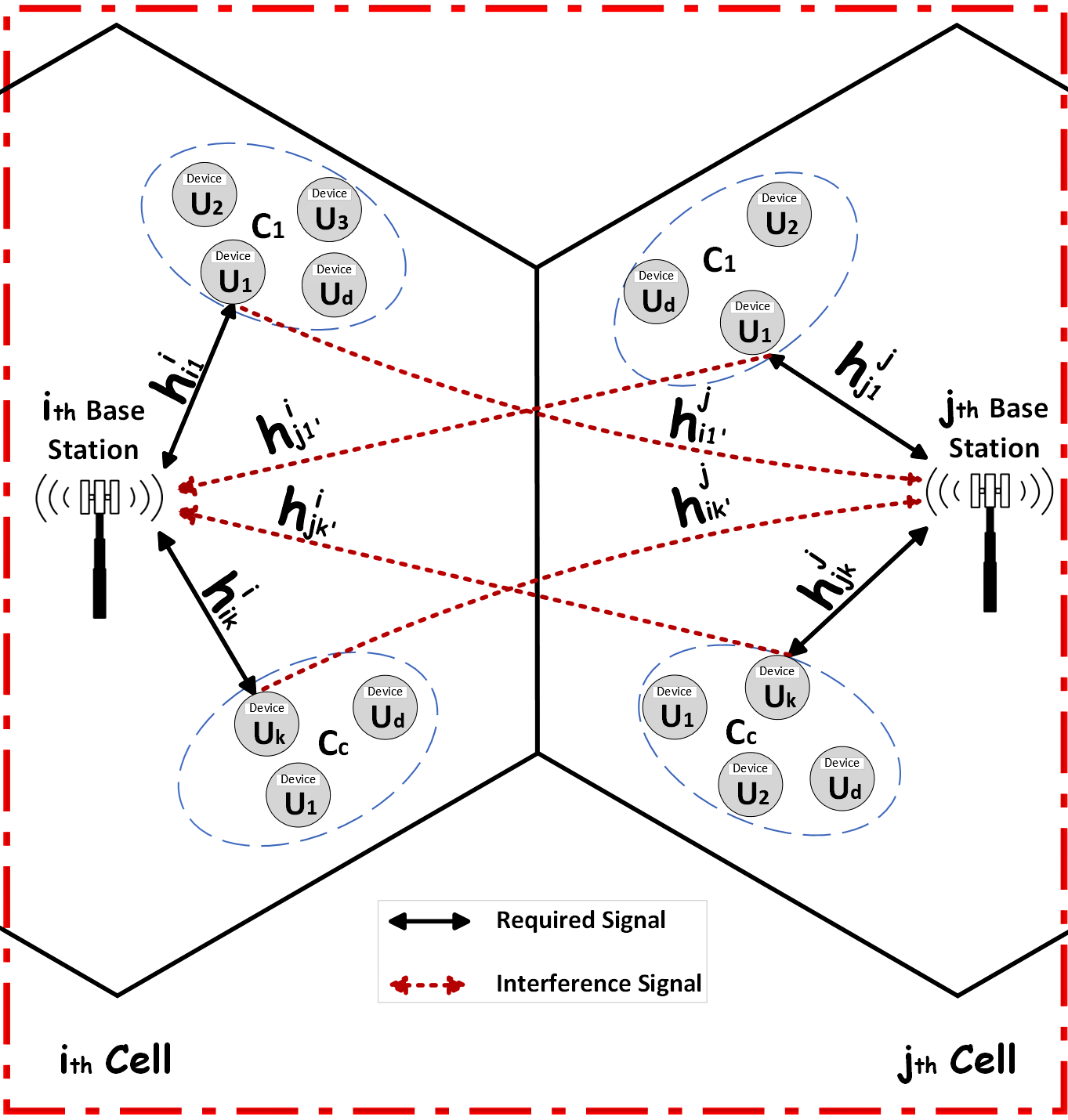}
    \caption{Required and interference signals of $i_{th}$ and $j_{th}$ cells}
    \label{fig:pilot}
\end{figure}

\subsection{Channel State Information}

\par The time-frequency is divided into coherence blocks $\tau_{c}$, and in each block, the channel response is frequency-flat and time-invariant. A portion of the coherence block, denoted as $\tau_{\rho}$, is allocated for transmitting a predefined pilot signal during uplink training. This allows the base station to acquire the channel state information of the $k_{th}$ device. The base station utilizes these pilot sequences to estimate the channels of $k_{th}$ device, represented by ${\hat {h}_{ik}^j} $, where $ {i,j \in \{1,...., L\}} $. The received pilot signals at $i_{th}$ base station can be written as,

\begin{equation}
{\mathbf Y{{'}_{i}}=  \underbrace {\sum _{k=1}^{K}\sqrt{{ \rho _{\mathrm{ul}}}} \mathbf h_{ik}^{i} \boldsymbol{\Phi }_{ik}^{H}}_{\mathrm {Desired~Pilots}} + \underbrace {\sum _{j=1,j\neq i}^{L} \sum _{k=1}^{K} \sqrt { { \rho _{\mathrm{ul}}}}  \mathbf h_{jk}^{i} \boldsymbol{\Phi }_{jk}^{H}}_{\mathrm {Inter\text{-}cell~PC}} + \underbrace {\mathbf n_{{i}}}_{\mathrm {AWGN}}}
\end{equation}

Where the pilot sequence of $k_{th}$ device of the $i_{th}$ cell is $\boldsymbol{\Phi }_{ik}^{H} \in \mathbb{C}^{\tau_{\rho}}$, and is transmitted with $\rho_{ul}$ power. The first term, desired pilots, means that the pilot sequences of devices of $i_{th}$ cell received at $i_{th}$ BS. However, the second term represents the inter-cell pilot contamination (PC) caused by the devices of $j_{th}$ cell, which use the same pilot sequence as of $i_{th}$ cell. The last term is additive white Gaussian noise $n_{{i}} \in \mathbb {C}^{M \times \tau_{\rho}} $ and is distributed as  $ \mathcal {N}_{\mathbb {C}} \left ({\mathbf {0}, \mathbf {\sigma}_{ul}^2 I_M }\right)$.

Generally, we assign an orthogonal pilot sequence to each device within a cell and reuse the same pilot sequences in other cells. Therefore,  $\boldsymbol{\Phi }_{ik} \boldsymbol{\Phi }_{jk'}^{H} = \tau_{\rho} $, if the device $k, k'$ are using the same pilot sequences in their respective $i_{th}$ cell and $j_{th}$ cell, otherwise, $\boldsymbol {\Phi }_{ik} \boldsymbol {\Phi }_{jk'}^{H} = 0 $.  Moreover, to perform the despreading operation, we multiply the received pilot signal with the known pilot sequence of the $k_{th}$ device.

\begin{align}
\mathbf y_{ik}&=\frac{1}{\tau_{\rho} \sqrt{\rho_{\mathrm{ul}}}}\mathbf Y'i\mathbf \Phi_{ik}^H &= \mathbf h_{ik}^i+\sum_{j=1, j\neq i}^L \mathbf h_{jk^{'}}^i+\frac{n_i}{\tau_{\rho}\sqrt{\rho_{\mathrm{ul}}}}
\end{align}

We are using a minimum mean squared error (MMSE) estimator to estimate channel ${  {h}}_{ik}^{i}$. The vector $\hat {  {h}}_{ik}^{i}$ that minimizes the mean squared error is,

\begin{align} \hat {  {h}}_{ik}^{i} =& {R}_{ik}^{i} \big ( \psi_{ik}^{i}  \big)^{-1} \mathbf{y}_{ik}=& {R}_{ik}^{i} \big ( {\sum _{j=1}^{L} } \mathbf {R}_{jk}^{i}+\frac {1}{\tau _{p}}\frac {\sigma _{\mathrm{ul}}^{2}}{{\rho _{\mathrm{ul}}}} \mathbf {I}_{M}  \big)^{-1} \mathbf{y}_{ik}\end{align}

Where the expected value of the channel estimation error is $ \Tilde{ h}_{ik}^{i} = { h}_{ik}^{i} - \hat {  {h}}_{ik}^{i} $ and can be written as,

\begin{align} \frac {\mathbb {E}\{ \| \mathbf {h}_{ik}^{i} - \hat { \mathbf {h}}_{ik}^{i} \|^{2} \}}{\mathbb {E}\{ \| \mathbf {h}_{ik}^{i} \|^{2} \}}=\frac { \mathrm {tr}{\big (\mathbf {C}_{ik}^{i}\big)}}{ \mathrm {tr}{\big (\mathbf {R}_{ik}^{i}\big)}}=1 -\frac { \mathrm {tr}{\big (\mathbf {R}_{ik}^{i} \big (\mathbf {\psi}_{ik}^{i}\big)^{-1} \mathbf {R}_{ik}^{i}\big)}}{ \mathrm {tr}{\big (\mathbf {R}_{ik}^{i}\big)}} \end{align}

The above-normalized minimum squared estimation error depends on PC, in which the pilot signals coming from other cells contaminate the channel estimates of $i_{th}$ cell, making it a limiting factor of a massive MIMO system.

\subsection{Signal Processing at Base Station}
Consider an uplink data transmission; the signal received at the $i_{th}$ base station is modeled as,    
\begin{equation} 
{\mathbf {Y}_{i} = \underbrace {  \sum _{k=1}^{K} \mathbf {h}_{ik}^{i} \mathbf s_{ik}}_{\mathrm {Desired~signals}} +  \underbrace { \sum _{j=1,j \neq i}^{L} \sum _{k=1}^{K} \mathbf {h}_{jk}^{i} \mathbf{s}_{jk}}_{\mathrm {Interference~Signal}} + \underbrace {\vphantom {\sum _{i=1,i\ne k}^{K} } \mathbf {n}_{i}}_{\mathrm {Noise}}}
\end{equation}

Where the $s_{ik}  \sim \mathcal {N}_{\mathbb {C}} \left ({ {0}, \mathbf {\rho}_{ul} }\right)$ represents the transmitted data signal by the $k_{th}$ device in the $i_{th}$ cell to the $i_{th}$ base station, where $\rho_{ul}$ is the uplink transmitted power. $n_{i}\sim \mathcal {N}_{\mathbb {C}}(0, \sigma_{ul}^2 I_{M})$ is an independent noise. 
To separate the required signal from the received signal in the ${i_{th}}$ cell, the base station utilizes the combining vector $\hat {W}^{i} = \mathbf{[\hat{w}_{i_{1}}, \hat{w}_{i_{2}},...... \hat{w}_{i_{k}} ]}$ to separate required signal from the interference and noise.

\begin{align*}
\hat {\mathbf w}^{H}_{ik} \mathbf Y_{i}=&\underbrace { \hat {\mathbf w}^{H}_{ik} \hat{\mathbf {h}}_{ik}^{i} \mathbf s_{ik}}_{\mathrm {Desired~Signal}} + \underbrace { \sum _{k'=1,k'\ne k}^{K} \hat{{\mathbf w}}^{H}_{ik} \hat{\mathbf {h}}_{ik'}^{i} s_{ik'}}_{\mathrm {Intra-cell~ Interference}} \\ &+ \underbrace {{\sum _{\substack {j=1, j\neq i}}^{L} \sum _{k=1}^{K}} \hat{{\mathbf w}}^{H}_{ik}  \hat{\mathbf h}_{jk}^{i} \mathbf s_{jk}}_{\mathrm {Inter-cell~ Interference}} + \underbrace {\hat{{\mathbf w}}^{H}_{ik}  \mathbf{n}_{ik}}_{\mathrm{Noise}} \tag {6}
\end{align*}

In the above equation, the desired signal is the required signal of the $k_{th}$ device and intra-cell interference is the added small portion of the signal from $k'$ devices within the same cell due to channel estimation error. However, inter-cell interference is the signal of the $k_{th}$ device of the $j_{th}$ cell using the same pilot sequence as of the $i_{th}$ cell. The spectral efficiency $\mathsf {SE}_{ik}$ can be computed as,

\begin{equation} { \mathsf {SE}}_{ik}= \frac {\tau _{up}}{\tau _{c}} \log _{2} (1 + {\gamma}_{ik})~\text {[bit/s/Hz]} 
 \tag {7} \end{equation}

Where $\tau _{up} = \tau _{c}-\tau _{\rho}$, represents uplink data samples and ${\gamma}_{ik}$ represents signal-to-interference and noise ratio (SINR). Therefore, the SINR of the $k_{th}$ device can be written as,

\begin{equation}
{{\gamma}_{ik} = \frac {\underbrace {{||\hat{{\mathbf w}}^{H}_{ik}} \hat{\mathbf h}_{ik}^{i}\rVert^2}_{Desired}} {\hat{{\mathbf w}}^{H}_{ik} \left (  \underbrace{ \sum _{k'=1, k'\ne k}^{K} || \hat{{\mathbf h}}_{ik'}^{i} ||^2   }_{intra-cell~ PC} +   \underbrace{{\sum _{\substack {j=1, j\neq i}}^{L} \sum _{k=1}^{K} ||\hat{{\mathbf h}}_{jk}^{i}||^2 }}_{inter-cell ~PC} + Z_{i} \right )\hat{{\mathbf w}}^{H}_{ik}}} \tag {8}
\end{equation}

Where 
\begin{equation} \mathbf {Z}_{i} = \sum \limits _{j=1}^{L} \sum \limits _{k=1}^{K} (\mathbf {R}_{jk}^{i} - \mathbf {R}_{jk}^{i} \big (\mathbf {\psi}_{jk}^{i}\big)^{-1} \mathbf {R}_{jk}^{i}) + \frac {\sigma _{\mathrm{ul}}^{2}}{\rho _{\mathrm{ul}}} \mathbf {I}_{M} \tag {9}   \end{equation}

Selecting a good receive combing vector would reduce the mean square error of the transmitted data signal $s_{ik}$ and the received signal at the base station. Therefore, this paper considers a multicell minimum mean squared error (M-MMSE) combining scheme. According to M-MMSE combining, ${\mathbf {W}^{i}_{\mathrm{M-MMSE}}}$ for $i_{th}$ cell can be written as, 

\begin{equation}
{\mathbf {W}^{i}_{\mathrm{M-MMSE}} = \left({\sum \limits _{j=1}^{L} \widehat { \mathbf {H}}_{j}^{i} {(\widehat { \mathbf {H}}_{j}^{i})}^{ {\mathrm {H}}} + \mathbf {Z}_{i} }\right)^{\!-1} \widehat { \mathbf {H}}_{i}^{i}} \tag {10}
\end{equation}

Where $ \hat{{H}}_{j}^{i} = \mathbf{[\hat{{h}}_{j1}^{i}, \hat{{h}}_{j2}^{i},..., \hat{{h}}_{jk}^{i} ]}$ is the combination of the individual channel of each device.

\section{Device Clustering}
Generally, an orthogonal pilot sequence is assigned to each device in a cell. During the uplink data transmission, a certain portion of the coherence block $\tau _{c}$ is dedicated to pilot symbols $\tau _{\rho}$. If there a $K$ devices within a cell, then at least we need $K$ orthogonal pilot symbols for each device for channel estimation. This means that by increasing the number of devices in a cell, we need more pilot symbols. As a result, we will have less space left in coherence blocks for data transmission, which is a limiting factor for enhancing spectral efficiency and scalability in a massive MIMO system.    
\par An inherent characteristic of IoT devices considered in our work is that they transmit data at a pre-defined regular interval (varies from millisecond to minutes) depending on the critical nature of the data.  Therefore, assigning an orthogonal pilot sequence to each device will not be viable. Thus, we propose a clustering algorithm to cluster devices that show a high spatial correlation between them. Moreover, we assigned an orthogonal pilot to each cluster, and within that cluster, each device will use the same pilot sequence in its turn while transmitting data. So, each device will take a certain data transmission time $T_{U_{d}}$, where ${d\in\{1,2,...,d\}}$ is the total number of devices in that cluster. This paper considers homogeneity in devices and data transmission size.  The case of non-homogeneous devices and variable data transmission sizes and rates will be addressed in our future work.

\subsection{Modified KFaster Medoid Clustering Algorithm}
 
\par In clustering, Kmedoid clustering is a popular approach because of its notable advantages. The primary benefits involve its decreased susceptibility to outliers and selecting a data point as the centroid that exhibits the closest distance to other nodes facilitates information transmission between cluster nodes. The prominent disadvantage is that the Kmedoid is computationally expensive. Schubert et al. \cite{b4} proposed a faster Kmedoid clustering algorithm which reduced the complexity by $\mathcal{O}(K(K-C))$. They introduced an eager swapping mechanism with different initializations and showed the proposed scheme produced good results even when eager swapping with random initialization was used. Therefore, this paper considers spatial correlation as a similarity metric for device clustering. Which can be calculated as,

\begin{align} S_{(U_{(ik)},U_{(ik')})}=\frac{\mathrm{Tr}\left({\mathbf R_{ik}^{i} \mathbf R^{iH}_{ik^{\prime }}}\right)}{\left|\left|\mathbf R_{ik}^{i}\right|\right|_F\left|\left|\mathbf R_{ik^{\prime }}^{i}\right|\right|_F},\;\; k\ne k^{\prime }, \tag {11} \end{align}

\par In this paper, we modified the Kfaster medoid clustering algorithm. We aim for a balanced device distribution in each cluster, allowing for slight variations. Fig. 2 compares the clustering algorithm with and without uniform distribution of devices within clusters. We can see that, till $K=100$, uniform and non-uniform distribution of clusters don't differ in omitted devices (Omitted devices are cluster devices unable to transmit due to time limitations or poor channel conditions). However, we can see a gradual increase in the number of omitted devices from $K=125$ onward. This means the difference could be higher if we go beyond $K=200$. Therefore, it is better to uniformly distribute devices in a cluster. 
\vspace{-0.2cm}

\begin{figure}[htp]
    \centering
    \includegraphics[width=8.5cm]{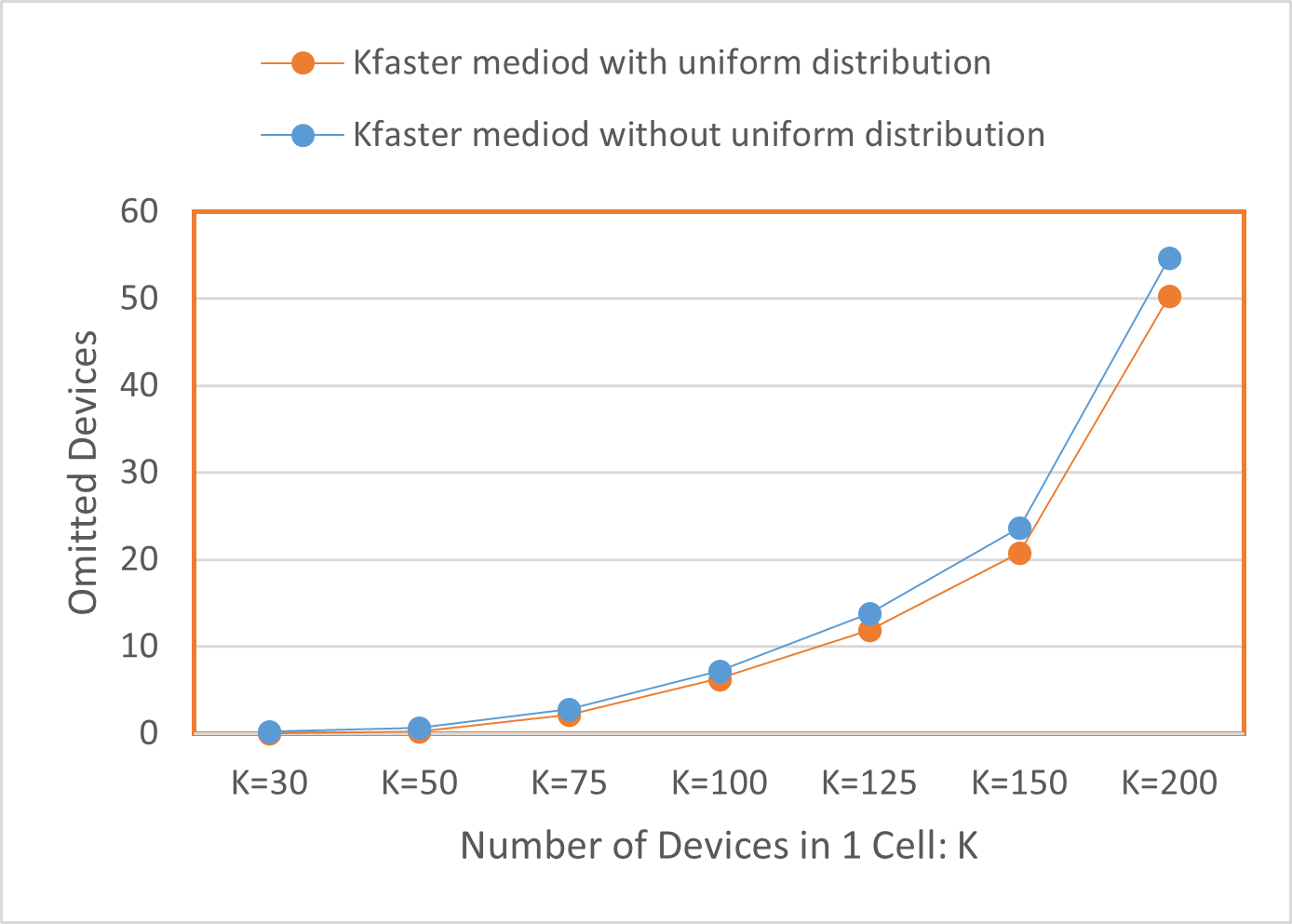}
        \vspace{-0.2cm}
    \caption{Comparison of clustered devices with and without uniform distribution}
    \label{fig:3}
\end{figure}
\vspace{-0.1cm}

\par In Algorithm 1, we initially define the inputs, which include the number of clusters: $C$, based on the available pilot sequence; spatial correlation matrix$:R$, to compute the similarity $S$ between devices; the set of devices$: {U_k: 1 \leq k \leq K}$, and the max cardinality of a cluster$:N$, which means that the number of devices in one cluster should not exceed $N$. The output is a set of clusters ${C \in {c_1, c_2, .....c_c}}$.

In line 1, the algorithm begins by randomly selecting $C$ nodes as medoids. In lines 2 and 3, it computes the nearest medoid $nearest(k)$, second nearest medoid $second(k)$, distance to nearest medoid $d_{nearest(k)}$, and distance to second nearest medoid $d_{second(k)}$ using the spatial correlation similarity matrix $S$. Between lines 4 to 7, If the closet medoid is $c_i$ for the $k_{th}$ device, and cardinality of ${c_i < N}$, then device $c_i$ is assigned  $k_{th}$ medoid otherwise it is assigned to the second nearest medoid. 

The proposed algorithm then uses a swapping technique to find a sub-optimal solution. From lines 12 to 22, the algorithm iterates through each device ${u_k}$ and computes the removal loss for each non-medoid ${u_o}$. In line 23, we pick a medoid having minimum removal loss. In line 24, we update the minimum deviation by adding the deviation of minimum removal loss with $u_k$. Between lines 25 to 29, we select the ${u_k}$ as a new medoid if the minimum change in deviation is less than zero. Furthermore, from lines 30 to 35, we again compute the $nearest(k)$, $second(k)$, $d_{nearest(k)}$, and $d_{second(k)}$ for each device, using the updated similarity matrix $S$ of new medoids and assigning devices to their respective clusters based on the same criterion as before. The algorithm returns the resulting set of clusters $C$ for $i_{th}$ cell and iterates for each cell.

\section{Weighted Graph Framework}

Once the nodes have been grouped into clusters using spatial correlation as a clustering criterion, the next step is to assign pilot sequences to the clusters. In this regard, pilot contamination restricts the performance of a massive MIMO system. One possible way to mitigate pilot contamination between two cells is to assign a pilot signal in such a way that minimizes the interference caused by a device in one cell over the device in another cell that is using the same pilot signal. Zeng et al. \cite{b5} proposed a graph coloring technique to mitigate pilot contamination in a cell-free massive MIMO system. However, the situation of a multi-cell massive MIMO system is different from a cell-free massive MIMO system. Therefore, we proposed a pilot assignment scheme for a multi-cell massive MIMO system to overcome pilot contamination. We reformulate the pilot assignment problem to the graph coloring problem and solve it using the maximum K-cut partitioning approach.     
\begin{algorithm}
\caption{Devices Clustering Algorithm}\label{alg:cap}
\textbf{Description:} Cluster devices based on spatial correlation \\
\textbf{Inputs:} ${R}$, $C$, $N$,
{$ U_k: 1 \leq k \leq K $ }

\textbf{Output:} Set of clusters $C\in \{c_{1},c_{2},...,c_{c} \}$.

\textbf{Modified Kfaster Medoid Clustering:}

\begin{algorithmic}[1]

\State Randomly pick nodes as medoids $m$: { $ m_i: 1 \leq i \leq C $ }

\State \textbf{foreach $u_k \notin m$ do}

\State  ~Using correlation similarity matrix $S$, Compute $~~~{nearest (k),~ second(k), ~d_{nearest}(k),~ d_{second}(k) }$
\State ~ \textbf{foreach $ m_i \in m $ do}
\State ~~ \textbf{if} $({nearest(k) == m_i~and~cardinality~of~ c_i<N })$
\State ~~~~~ $c_i \gets u_i$
\State ~~ \textbf{else:} $c_{second(k)} \gets u_i$  
\State \textbf{Swapping to Find Sub-Optimal Solution}

\State $u_{\text{last}} \gets \text{invalid}$
\State $ \Delta TD_{m_1}, \dots, \Delta TD_{m_c} \gets$ Initial Removal Loss
\Repeat
\State \textbf{foreach $u_k \notin m$ do}

\State ~~\textbf{break outer loop} \textbf{if} $u_k == u_{\text{last}}$
\State ~~$\Delta TD \gets \Delta TD_{m_1}, \dots, \Delta TD_{m_c} $ 
\State ~~$\Delta TD_{u_k} \gets 0$
\State ~~{\textbf{foreach~} $u_o \notin m$,}
\State ~~~~$d_{oj} \gets d(u_o, u_k)$
\State ~~~~\textbf{If~}{$d_{oj} < d_{nearest}(k)$} 
\State ~~~~~~$\Delta TD_{u_{k}} \gets \Delta TD_{u_{k}}  + d_{oj} - d_{nearest}(k)$
\State ~~~~~~~~$\Delta TD_{nearest(k)} \gets \Delta TD_{nearest(k)} + d_{nearest(k)} - d_{second(k)}$
\State ~~~~$ \textbf{Else~If~}{d_{oj} < d_{second_{(k)}}}$
\State ~~~~$\Delta TD_{nearest(k)} \gets \Delta TD_{nearest(k)} + d_{oj} - d_{second}(k)$
\State ~~$min \gets \text{arg min } \Delta TD_{i}$
\State ~~$\Delta TD_{min} \gets \Delta TD_{min} + \Delta TD_{u_k}$ 
\State ~~\textbf{If~}{$\Delta TD_{min} < 0$} 
\State ~~~~swap roles of medoid $m^*$ and non-medoid $u_k$
\State ~~~~$TD \gets TD + \Delta TD_{min}$
\State ~~~~update $\Delta TD_{m_1}, \dots, \Delta TD_{m_c} $ 

\State ~~$u_{\text{last}} \gets u_k$

\State \textbf{foreach $u_k \notin m$ do}

\State ~Using updated similarity matrix $S$, Compute $~~~~~~~~~~~{nearest (k),~ second(k), ~d_{nearest}(k),~ d_{second}(k) }$
\State ~ \textbf{foreach $ m_i \in m $ do}
\State ~~~~ \textbf{if} $({nearest(k) == m_i~\&~cardinality~of~c_i<N })$
\State ~~~~~~ $c_i \gets u_i$
\State ~~~ \textbf{else:} $c_{second(k)} \gets u_i$ 
\Until{convergence}
\State \textbf{return} $C$

\end{algorithmic}
\end{algorithm}

\begin{equation*}
I_{c,c'} = \sum\limits_{\substack{{c}=1}}^{C}\sum\limits_{\substack{{c'}=c}}^{C}\left|\frac{{\sum\limits_{\substack{\mathrm{k \in c_{c'}}}}\mathrm{\beta}_{jk}^i}/{|c_{c'}|}}{{\sum\limits_{\substack{\mathrm{k \in c_c}}}\mathrm{\beta}_{ik}^i}/{|c_c|}} + \frac{{\sum\limits_{\substack{\mathrm{k \in c_c}}}\mathrm{\beta}_{ik}^j}/{|c_c|}}{{\sum\limits_{\substack{\mathrm{k \in c_{c'}}}}\mathrm{\beta}_{jk}^j}/{|c_{c'}|}}\right| \tag {12}
\end{equation*}

In the above equation, $I_{c,c'}$ is the sum of the ratio of average interference caused by $c_c \gets c_{c'}$ and $c_{c'} \gets c_c$, where $c_{c'}~and~c_c$ are two clusters in two different cells, which can be using same pilot sequence. To combine the impact of multiple devices within a cluster, we take the average $\beta$ of devices within a cluster, where $|c_c|$ represents the cardinality of cluster $c_c$. In the above equation, $\beta$ is a large-scale fading coefficient containing geometric attenuation and shadow fading.

\begin{algorithm}

\caption{Pilot Assignment Algorithm}\label{alg:pilot_assignment}
\textbf{Description:} This algorithm minimizes inter-cell PC.\\
\textbf{Inputs:}
 $ L,K, C, \beta_{ik}^j $

\textbf{Output:} Set of Sub-Graphs $V\in \{V_{1},V_{2},.....V_{c} \}$.

\textbf{Algorithm:}

\begin{algorithmic}[1]
\State Unassigned clusters of $i_{th}$ cell ${P_{ic}: 1\leq c\leq C , 1\leq i\leq L}$
\State Subgraphs of cluster set $V_s = \varnothing$, \textbf{for} $1\leq s\leq \tau_{\rho}$
\State \textbf{For~}{$1\leq s\leq \tau_{\rho}$}
\State ~Arbitrarily select $i_{th}$ cell $c_{c}$, $V_s=V_s\cup c_i$; $P_{\text{ic}}=P_{\text{ic}}-{c_{i}}$

\State \textbf{For} {$ 1\leq l \leq L $}
\State ~~${I_{c,c'}=0}$
\State ~~\textbf{For} {$ 1\leq j \leq L $}
\State ~~~~\textbf{If} {$Assigned~Neighbors~Cells~(i,j)==1$  }
\State ~~~~~~\textbf{For}{~$1\leq c< c'\leq C$}
\State ~~~~~~~~${I_{c,c'}~ += \frac{{\sum\limits_{\substack{\mathrm{k \in c_{c'}}}}\mathrm{\beta}_{jk}^i}/{|c_{c'}|}}{{\sum\limits_{\substack{\mathrm{k \in c_c}}}\mathrm{\beta}_{ik}^i}/{|c_c|}} + \frac{{\sum\limits_{\substack{\mathrm{k \in c_c}}}\mathrm{\beta}_{ik}^j}/{|c_c|}}{{\sum\limits_{\substack{\mathrm{k \in c_{c'}}}}\mathrm{\beta}_{jk}^j}/{|c_{c'}|}}}$

\State ~~ max indices $\gets$ descend-sorted~($I_{c,c'}$)
\State ~~ \textbf{Foreach $ max~index \in max~indices$ do}
\State ~~~~~ ${V_{min~interference} \gets V_{min~interference} + P_{i_{max~index}} }$
\State ~~~~~ ${P_{ic}  \gets P_{ic} - P_{i_{max~index}}  }$

\State Update and return subgrahps $V_s$ for $1\leq s\leq \tau_{\rho}$
\end{algorithmic}

\end{algorithm}

\par In Algorithm 2, from lines 1 to 4, we randomly assign clusters of the first cell to subgroups $V_s$. Between lines 5 to 10, we compute the interference matrix for $i_{th}$ cell using neighboring cells' already assigned pilot assignments. In line 11, we sorted the indices of $i_{th}$ cell clusters with interference from all neighboring cells in descending order. Afterward, from lines 12 to 14, we pick the maximum interference cluster of $i_{th}$ cell and assign it to a subgroup having the least interference to $i_{th}$ cell from the neighboring cell, and we keep iterating this procedure until we assign all the highest interference clusters to the least interference-causing subgroups for all cells.

\section{Simulation Results}
\par We performed simulations of the proposed technique using Matlab and associated Communication Tool Box. We consider a hexagonal network of 16 cells, utilizing the wrap-around technique to avoid the boundary effect at the edges. Each cell has $K$ uniformly distributed devices with a minimum distance to the serving base station of 35m. We are considering 50 random channel realizations. We are allocating orthogonal pilot sequences to each cluster within a cell and reusing them in all other cells. Therefore, there are $C$ orthogonal pilot signals assigned to each cell. Further details on the parameters used in this paper are in Table 1.

\begin{table}[htbp]
\caption{Simulation Parameters}
\begin{center}
\vspace{-0.5cm}
\begin{tabular}{|c|c|c|c|}
\hline

\textbf{Parameter} & \textbf{Value}& \textbf{Parameter}& \textbf{Value} \\

\hline
Cell Area & $0.25\times0.25km^2$ & Channel gain:1 km & -148.1 dB \\
\hline
Pathloss exponent & $\alpha = 3.76$ & Shadow fading & $\sigma_{sf}=10$ \\
\hline
Receiver noise & $-94 dBm$ & UL transmit power & 20 dBm \\
\hline

Coherence Block & 200 sample & M & 64 \\
\hline

Data Size & 500 bytes  & bandwidth & 12.5 Khz \\
\hline

\end{tabular}
\label{tab1}
\end{center}
\end{table}

\par This paper considers a usecase of demand response (DR) management in a smart grid. DR effectively regulates and allocates resources to maintain a steady flow of smart grids but requires improved communication technologies \cite{b6}. However, 5G-enabled IoT devices enhance DR management by real-time updates of smart meters' energy consumption data to energy providers. We consider DR due to its significant predictive capability, with only a portion of devices having negligible impact of omitted devices.

This paper considers smart meters as IoT devices that transmit energy consumption data to the base station after regular intervals. We are considering a narrow bandwidth of 12.5 KHz in the 902 MHz ISM band. In Fig 3, we compare the impact of the increasing number of devices $K$ and available orthogonal pilot signals/clusters $C$ over the omitted devices. Omitted devices are those devices that cannot transmit data due to bad channel conditions, limited orthogonal pilot sequences, or time-constrained, which is 1 second in the first case.

\begin{figure}[htp]
    \centering
    \includegraphics[width=8.5cm]{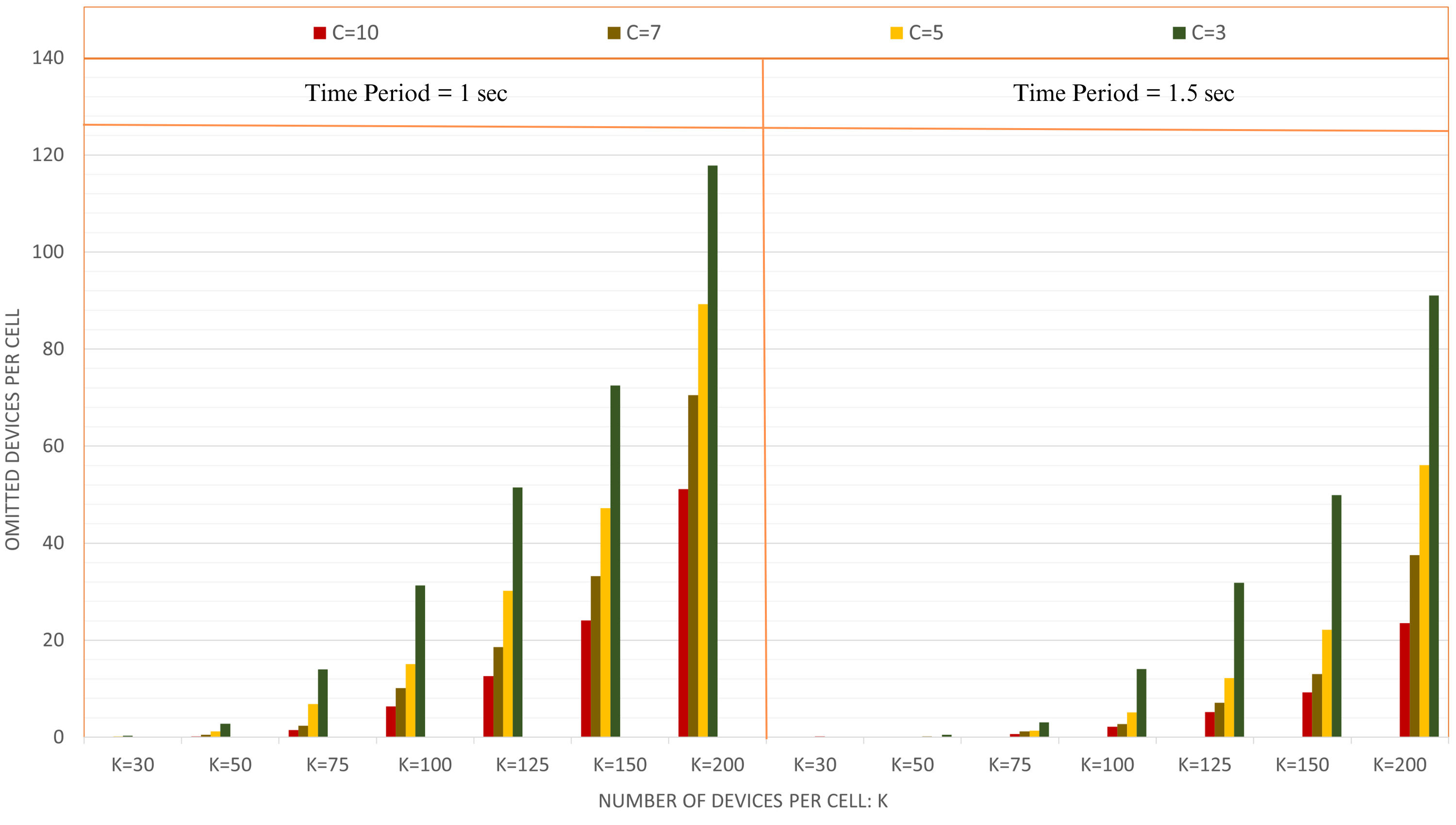}
        \vspace{-0.2cm}
    \caption{Impact of increasing devices in a cell over omitted devices}
    \label{fig:1}
\end{figure}

\par Fig. 3, we can see that as the number of available orthogonal pilot signals/cluster $C$ increases, the number of omitted devices decreases exponentially for a fixed number of devices $K$. However, initially omitted devices are close to zero till $K=50$ and linearly increase with the number of devices. Moreover, when the time period of smart meters increases by 50 percent (1.5 seconds), this results in reducing the omitted devices by nearly 40 percent, as shown in Fig. 3. Therefore, with ten orthogonal pilot sequences per cell, we can accommodate 175 IoT devices, with just 25 omitted devices when $K=200$.  

\par Fig. 4 depicts the spectral efficiency (SE) achieved with and without clustering in any given cell, assuming different number of devices ($K$) in the cell. The coherence block length is fixed to be 200 samples per communication round. Based on the simulation parameters listed in Table 1, our results show that the SE of IoT devices transmitting without clustering initially increases until $K$=46 and reaches a maximum of 89.1 $(bits\diagup s\diagup Hz\diagup  cell)$. It starts decreasing due to the fact that without clustering, with the increase in the number of devices in a cell, most part of the coherence block gets consumed by the transmission of pilot samples rather than data transmission. On the other hand, assuming all simulation parameters remain unchanged, the SE of the clustering-based scheme coupled with graph-coloring-based pilot sequence assignment, introduced in this paper, remains stable around a constant value of 81 $(bits\diagup s\diagup  Hz\diagup  cell)$ even when the maximum number of devices ($K=200$) in a cell is reached for $C=25$ clusters with 8.17\% omission rate. It is due to the fact that with clustering and conflict-free pilot sequence assignment using graph-coloring, we acquire the ability to schedule a fixed number of devices transmitting in any given round without channel collisions. 
    \vspace{-0.2cm}

\begin{figure}[htp]
    \centering
    \includegraphics[width=8.5cm]{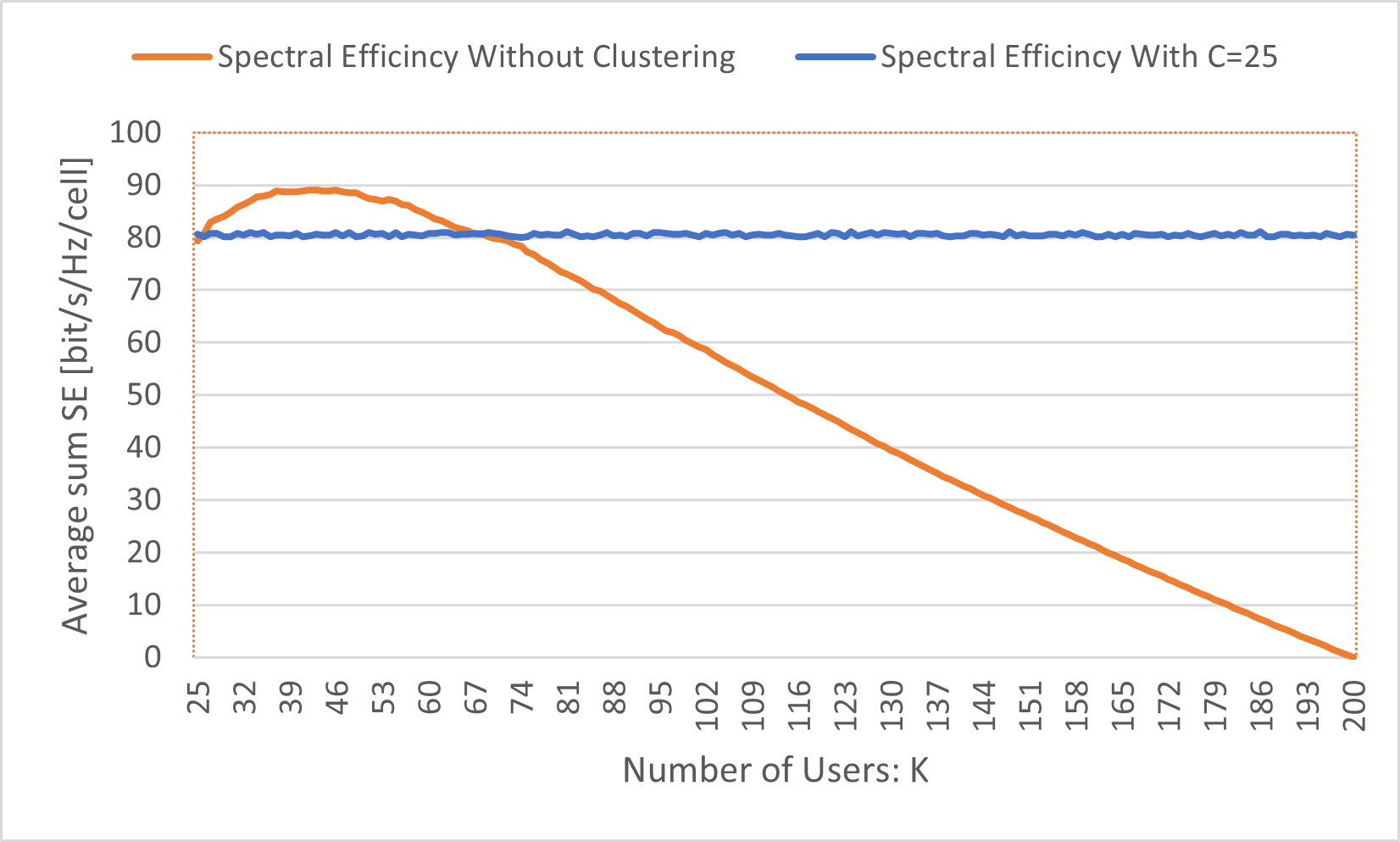}
    \vspace{-0.2cm}
    \caption{Impact of average SE by increasing users with fixed coherence block}
    \label{fig:2}
\end{figure}
    \vspace{-0.2cm}

\section{Discussion and Future Research Directions }
In this paper, we addressed the issue of scalability of massive MIMO systems by considering the orthogonal pilot signal assignment problem as a limiting factor in integrating a large number of IoT devices. Orthogonal pilot signals are scarce and should be assigned effectively to fulfill the orthogonality requirement within a cell to reduce intra-cell pilot contamination. Therefore, we proposed that instead of assigning an orthogonal pilot signal to each device, we can assign one pilot signal to each group, and each device can take its turn to transmit data.
\par Figure 4 results demonstrate that assigning an orthogonal pilot sequence to each device is expensive and inefficient. Furthermore, our approach enables the integration of numerous devices using minimal orthogonal pilot sequences. This reduces pilot symbols, conserves coherence block space for signal transmission, and enhances spectral efficiency.       

\par In our future work, we aim to consider minimizing the number of omitted devices. Moreover, prioritizing certain devices over others in a cluster will help us to minimize omitted devices. Therefore, a mechanism is needed to prioritize devices that exhibit better channel conditions over others in a cluster.       
\par In our current work we have considered homogeneous devices, i.e., they are identical or very similar in terms of their specifications, features, or functionality. Incorporating heterogeneous devices, particularly in terms of different sample rate, sample size, and the time-critical nature of the data, pose significant additional optimization constraints that must be addressed. Furthermore, an intelligent mechanism would be needed to assign devices to different clusters while keeping in view the specification and critical nature of each device's data.

\section{Conclusion}

Massive MIMO is crucial for 5G networks, enabling higher spectral efficiency, coverage, and energy efficiency required for IoT deployment. Acquisition of channel state information using orthogonal pilot signals plays a key role in massive MIMO systems. Due to scarce resources of orthogonal pilot signals and ineffective pilot assignment limits the scalability and spectral efficiency of massive MIMO systems. To address this issue, we proposed a novel pilot allocation scheme based on data transfer patterns from IoT devices and, assigned orthogonal pilot sequences to clusters of devices instead of individual devices and minimized pilot contamination using the max k-cut graph partitioning problem. Our simulation results showed the effectiveness of the proposed scheme in incorporating a large number of devices with a few orthogonal pilot sequences without sacrificing spectrum usage efficiency.  Possible future research directions are proposing advanced methods for pilot contamination, prioritizing certain devices based on their channel quality, and intelligent mechanisms for incorporating heterogeneous devices in a smarter way.


\begin{thebibliography}{00}


\bibitem{b3} 
C. Feres and Z. Ding, "An Unsupervised Learning Paradigm for User Scheduling in Large Scale Multi-Antenna Systems," in IEEE Transactions on Wireless Communications, vol. 22, no. 5, pp. 2932-2945, May 2023, doi: 10.1109/TWC.2022.3215471.

\bibitem{b1} L. Sanguinetti, E. Björnson and J. Hoydis, "Toward Massive MIMO 2.0: Understanding Spatial Correlation, Interference Suppression, and Pilot Contamination," in IEEE Transactions on Communications, vol. 68, no. 1, pp. 232-257, Jan. 2020, doi: 10.1109/TCOMM.2019.2945792.


\bibitem{b2}
Emil Björnson, Jakob Hoydis and Luca Sanguinetti (2017), “Massive MIMO Networks: Spectral, Energy, and Hardware Efficiency,” Foundations and Trends® in Signal Processing: Vol. 11: No. 3-4, pp 154-655. http://dx.doi.org/10.1561/2000000093




\bibitem{b4}  Schubert, E., and Rousseeuw, P. J. (2021). Fast and eager k-medoids clustering: O(k) runtime improvement of the PAM, CLARA, and CLARANS algorithms. In 2021 IEEE International Conference on Big Data (Big Data) (pp. 1-6). IEEE.

\bibitem{b5}W. Zeng, Y. He, B. Li and S. Wang, "Pilot Assignment for Cell Free Massive MIMO Systems Using a Weighted Graphic Framework," in IEEE Transactions on Vehicular Technology, vol. 70, no. 6, pp. 6190-6194, June 2021, doi: 10.1109/TVT.2021.3076440.


\bibitem{b6} Hui, H., Ding, Y., Shi, Q., Li, F., Song, Y., and Yan, J. (2020). 5G network-based Internet of Things for demand response in smart grid: A survey on application potential. Applied Energy, 257, 113972.



\end{thebibliography}
\end{document}